\documentclass[twocolumn,pra,twocolumn,superscriptaddress]{revtex4-2}
\usepackage[english]{babel}
\usepackage[utf8]{inputenc}
\usepackage[colorinlistoftodos, color=green!40, prependcaption]{todonotes}
\usepackage{amsthm} 
\usepackage{mathtools}
\usepackage{amsmath}
\usepackage{physics}
\usepackage{xcolor}
\usepackage{graphicx}
\usepackage{mathrsfs}
\usepackage{mathbbol}
\usepackage[pdftex, pdftitle={Article}, pdfauthor={Author}]{hyperref} 
\DeclareMathOperator*{\argmin}{arg\,min}
\usepackage{color,soul}

\begin{document}

\title{On Circuit Depth Scaling For Quantum Approximate Optimization}

\author{V.~Akshay}
\email{akshay.vishwanathan@skoltech.ru}
    \affiliation{Skolkovo Institute of Science and Technology, 30 Bolshoy Boulevard, Moscow 121205, Russian Federation}

\author{H.~Philathong}
    \affiliation{Skolkovo Institute of Science and Technology, 30 Bolshoy Boulevard, Moscow 121205, Russian Federation}
    
\author{E.~Campos}
    \affiliation{Skolkovo Institute of Science and Technology, 30 Bolshoy Boulevard, Moscow 121205, Russian Federation}
    
\author{D.~Rabinovich}
    \affiliation{Skolkovo Institute of Science and Technology, 30 Bolshoy Boulevard, Moscow 121205, Russian Federation}

\author{I.~Zacharov}
    \affiliation{Skolkovo Institute of Science and Technology, 30 Bolshoy Boulevard, Moscow 121205, Russian Federation}
\author{Xiao-Ming Zhang}
    \affiliation{Department of Physics, City University of Hong Kong, Tat Chee Avenue, Kowloon, Hong Kong SAR, China} 
\author{J.D.~Biamonte}
    \affiliation{Skolkovo Institute of Science and Technology, 30 Bolshoy Boulevard, Moscow 121205, Russian Federation}

\date{\today} 

\begin{abstract}
Variational quantum algorithms are the centerpiece of modern quantum programming. These algorithms involve training parameterized quantum circuits using a classical co-processor, an approach adapted partly from classical machine learning. An important subclass of these algorithms, designed for combinatorial optimization on currrent quantum hardware, is the quantum approximate optimization algorithm (QAOA). Despite efforts to realize deeper circuits, experimental state of the art implementations are limited to fixed depth. However, it is known that {\it problem density}---a problem constraint to variable ratio---induces under-parametrization in fixed depth QAOA. Density dependent performance has been reported in the literature, yet the circuit depth required to achieve fixed performance (henceforth called critical depth) remained unknown. Here, we propose a predictive model, based on a \textit{logistic saturation conjecture} for critical depth scaling with respect to density.
\noindent Focusing on random instances of MAX-2-SAT, we test our predictive model against simulated data with up to 15 qubits. We report the average critical depth, required to attain a success probability of 0.7, saturates at a value of 10 for densities beyond 4. We observe the predictive model to describe the simulated data within a $3\sigma$ confidence interval. Furthermore, based on the model, a linear trend for the critical depth with respect problem size is recovered for the range of 5 to 15 qubits.

\end{abstract}

\maketitle


\section{Introduction}

The quantum approximate optimization algorithm (QAOA) was introduced to solve combinatorial optimization problems by Farhi \textit{et al.}~\cite{farhi2014quantum}. In this algorithm, the candidate solution is prepared using a fixed ansatz structure with tunable parameters that are variationally adjusted to minimize an objective function. QAOA was inspired by adiabatic quantum computation \cite{kadowaki1998quantum,farhi2001quantum,boixo2014evidence}, where a system prepared in the ground state of an initial Hamiltonian evolves towards the ground state of a problem Hamiltonian, given a slow enough evolution. The circuit structure of QAOA shares resemblance to adiabatic evolution via a finite depth Trotterization procedure, and thus allows for recovering the exact ground state of the problem Hamiltoninan in the infinite depth limit. In contrast, a fixed $p$--depth QAOA circuit, consists of alternating applications of the complex exponentiation of the problem Hamiltonian and a so called mixer Hamiltonian, accounting for $2p$ variational parameters.

Although QAOA recovers adiabatic evolution in the infinite depth limit, short depth circuits have indeed been shown to offer some benefits in approximating solutions to low density instances of combinatorial optimization problems \cite{farhi2014quantum,farhi2014quantumBOCP,wang2018quantum}. Farhi \textit{et al.}~\cite{farhi2014quantum} studied the performance of QAOA applied to the MAX-CUT problem on 3-regular graphs and reported that for $p=1$, QAOA guarantees a cut that is at least $0.6924$ of the optimal cut. Furthermore, the quality of approximation is shown to improve with increasing $p$. Beyond combinatorial optimization problems, Farhi and Harrow \cite{farhi2016quantum} reported that the split operator structure of QAOA might be a suitable candidate for establishing quantum supremacy, since the output distributions of such circuits may not be efficiently sampled classically. Moreover, in the works of Jiang \textit{et al.}~\cite{jiang2017near} and Morales \textit{ et al.}~\cite{morales2018variational}, the $O(\sqrt{N})$ query complexity of Grover’s algorithm is recovered by a class of QAOA circuits, thus establishing quantum advantage within the QAOA framework. These along with universality results \cite{biamonte2021universal,morales2020universality,lloyd2018quantum} make QAOA a promising model for NISQ era devices. 

Although several advantages have been reported  \cite{farhi2014quantum,farhi2014quantumBOCP,wang2018quantum,jiang2017near,morales2018variational,anschuetz2019variational,ho2019efficient,farhi2016quantum, yang2017optimizing}, effects limiting the performance of QAOA  have also been discovered \cite{akshay2020reachability, akshay2021reachability, campos2021training}. Considering the energy error in approximation as a performance metric, Akshay \textit{et al.}~\cite{akshay2020reachability, akshay2021reachability} demonstrated \textit{reachability deficits}, an effect limiting performance for fixed depth QAOA on high density problem instances. Deeper circuits therefore become a necessity and hence come with overheads in classical outerloop optimization. While approaches such as layerwise training aim to reduce this classical computational cost, Campos \textit{et al.}~\cite{campos2021training} showed that such a strategy would not improve performance beyond some threshold depth. Other approaches aimed at reducing the classical computation cost exploit what are known as \textit{concentration effects} \cite{Farhi2019a,Streif2020, akshay2021parameter,wecker2016training}. Although these effects allows one to heuristically guess near-optimal parameters at fixed depth and for increasing number of qubits, they fail to address the depth required to guarantee fixed performance.

Here we propose a {\it logistic saturation conjecture} which states that the 
minimum depth required for QAOA to achieve fixed performance up to some tolerance (henceforth called critical depth) scales logistically with problem density. We numerically test this conjecture for the case of MAX-2-SAT with up to $15$ qubits and maximum density of $4$, and observe (i) the predictive model describes the simulated data within a $3\sigma$ confidence interval and (ii) the critical depth scales linearly with problem size in this dataset. 

\section{Background}

\subsection{Quantum Approximate Optimization Algorithm}
The objective function of a combinatorial optimization problem is prescribed as a map from the assignment of variables to non-negative real numbers $\mathcal{C}:\{0,1\}^{\times n} \rightarrow \mathbb{R_+}$. Under a vector space embedding \cite{biamonte2008nonperturbative, lucas2014ising}, the objective function then induces a corresponding problem Hamiltonian $\mathcal{H}_c$, which encodes the solutions to the problem instance in its ground state space,
\begin{equation}
    \mathcal{H}_c = \sum_{z \in \{0,1\}^{\times n}} \mathcal{C}(z) \ketbra{z}{z}, 
\end{equation} where $\ket{z}$ are the computational basis vectors in the $2^n$ dimensional Hilbert space of $n$ qubits. 

The optimization problem aims to find assignments such that the objective function $\mathcal{C}(z)$ is minimized. This is equivalent to finding a state $\ket{z^*}$ such that
\begin{equation}
    \ket{z^*} \in \text{span} \lbrace \underset{z \in \{0,1\}^{\times n}}{\argmin} \bra{z}\mathcal{H}_c\ket{z} \rbrace.
\end{equation} 
Measuring the state $\ket{z^*}$ in the computational basis outputs a bitstring which is a solution to the optimization problem, $\mathcal{C}$.

To approximate solutions to an optimization problem instance using a $p$-depth QAOA circuit, we generate ansatz states $\ket{\psi(\boldsymbol{\gamma},\boldsymbol{\beta})}$ on a quantum computer with tunable real parameters $\boldsymbol{\gamma}, \boldsymbol{\beta} \in [0,2\pi)^{\times p} , [0, \pi)^{\times p}$. These states are prepared by applying an alternating sequence of $2p$ parameterized unitary gates on the initial state $\ket{+}^{\otimes{n}} = \frac{1}{\sqrt{2^n}} \sum_{z \in \{0,1\}^{\times n}} \ket{z}$ as,
\begin{equation}\label{qaoa equation}
        \ket{\psi(\boldsymbol{\gamma},\boldsymbol{\beta})}=\prod_{k=1}^{p} e^{-i \beta_{k}\mathcal{H}_{x}} \cdot e^{-i \gamma_{k}\mathcal{H}_c}\ket{+}^{\otimes{n}},
\end{equation} with the mixing Hamiltonian, $\mathcal{H}_{x}=\sum_{l=1}^n X_{l}$. Here $X_l$ is the Pauli matrix applied to the $l^{th}$ qubit.

The ansatz state in Eq.~\eqref{qaoa equation} is iteratively adjusted via an outer-loop optimization routine to obtain an energy approximation as: 

\begin{equation}
    \min_{\boldsymbol{\gamma},\boldsymbol{\beta}}~ \bra{\psi(\boldsymbol{\gamma},\boldsymbol{\beta})}\mathcal{H}_c\ket{\psi(\boldsymbol{\gamma},\boldsymbol{\beta})} \geq \min\left(\mathcal{H}_c\right).
    \label{minimization}
\end{equation}

Algorithmic performance of QAOA can be assessed under performance metrics such as (i) energy error in approximation and (ii) ground state overlap \cite{akshay2020reachability,akshay2021reachability,campos2021training}. Both these quantifiers feature shortcomings as energy approximations fail to indicate closeness to optimal solutions while computing ground state overlap requires prior knowledge of optimal solutions. However, it can be demonstrated that both metrics inter-relate and therefore have some advantages as a theoretical tool for studying algorithmic performance.    

Let $\mathcal{H}^{\dagger}=\mathcal{H} \geq 0$ be an optimization problem instance with $d$-degenerate ground state energy $\lambda_0$. In its eigenbasis 
$\ket{\phi_{j}}$, we can write $\mathcal{H} = \lambda_0 \sum_{j=1}^{d} \ketbra{\phi_j}{\phi_j} + \sum_{k > d} \lambda_k  \ketbra{\phi_k}{\phi_k}$, with subsequent excited state energies, $\lambda_0 < \lambda_k \leq \lambda_{max} $ for $k > d$, separated by a gap $\Delta = \min_{k > d} |\lambda_0 - \lambda_k|$. For any arbitrary state $\ket{\psi}$, the ground state overlap is calculated as:  

\begin{equation} \label{overlap}
    g(\psi) =  \sum_{j=1}^{d} \abs{\braket{\phi_{j}}{\psi}}^{2}.
\end{equation}

It is straightforward now to derive (see Appendix \ref{Appx:stability}) the bounds relating the energy error in approximations $\bra{\psi}\mathcal{H}\ket{\psi} - \lambda_0$, and the ground state overlap $\text{g}(\psi)$:

\begin{equation} \label{energy-ovelap}
    1 - \frac{\bra{\psi}\mathcal{H}\ket{\psi} - \lambda_0}{\Delta} \leq  g(\psi) \leq 1 - \frac{\bra{\psi}\mathcal{H}\ket{\psi} - \lambda_0}{\lambda_{max}}.
\end{equation}  Note that the relation works when the error in approximation is less than the spectral gap $\Delta$. These bounds are the extended version of the variational stability lemma in \cite{biamonte2021universal} to the degenerate case and allows one to recover an estimate on the ground state overlap from the measured energy.

\subsection{Satisfiability}

Boolean satisfiability or SAT, is the problem of determining satisfiability of Boolean formulae. More specifically, one aims to decide whether a given formula can be satisfied by assigning truth values to the variables. Limited to $k$-variables or literals per clause, and expressed in the conjunctive normal form (CNF), $k$-SAT is \textbf{NP}-complete for $k \geq 3$ \cite{cook1971complexity}. 
$k$-SAT clauses are randomly generated to form instances by uniformly selecting unique $k$-tuples from the union of a variable set (cardinality $n>k$) and its element wise negation. Generated random instances admit an order parameter called clause or problem density given by,  
\begin{equation}
    \alpha=m/n,
\end{equation} 
where $m$ is the number of clauses and $n$ is the number of Boolean variables in the $k$-SAT instance.

The clause density plays an important role in the study of satisfiability phase transitions. It has been observed that for random $k$-SAT instances, an abrupt change in satisfiability occurs at some critical clause density, $\alpha_c$. Though this criticality has been studied only empirically for $3$-SAT ($\alpha_c \approx 4.27$), it is proven for $2$-SAT \cite{goerdt1996threshold}. Furthermore, it is observed that the computational resources required for SAT solvers increase at the transition point with an easy-hard-easy pattern. This behaviour suggest that the hard instances concentrate around the critical clause density \cite{crawford1996experimental,friedgut1999sharp,selman1996critical,selman1996generating,chvatal1992mick,goerdt1996threshold}.

In this work we focus on the optimization version of $k$-SAT, otherwise called MAX-$k$-SAT. MAX-$k$-SAT is the canonical {\bf NP}-hard optimization problem (for $k\geq2$) \cite{barahona1982computational} where one seeks to determine variable assignments that maximize the number of satisfied clauses in a given instance. Similar to decision $k$-SAT, MAX-$k$-SAT also features a phase transition with an easy-hard pattern \cite{zhang2001phase}. Although the point of criticality differs for the optimization and decision variants of these problems, they coincide for $k = 2$ with $\alpha_c = 1$ \cite{coppersmith2004random}. 

With standard Ising embedding techniques \cite{biamonte2008nonperturbative,lucas2014ising}, MAX-2-SAT instances can be mapped onto $2$-local Ising Hamiltonians of the form:

\begin{equation} \label{sathamiltonian}
\mathcal{H} = \sum_{i<j} \mathcal{J}_{ij} Z_{i}Z_{j} + \sum_i h_{i} Z_{i},
\end{equation}
with appropriate coefficients and scale such that: (i) for satisfiable assignments, $\ket{z}$, $\bra{z}\mathcal{H}\ket{z} = 0$ and (ii)
for un-satisfiable assignments $\bra{z}\mathcal{H}\ket{z} = m'$, where $m'$ is the number of clauses violated by the assignment. A detailed construction of such Hamiltonians is shown in Appendix \ref{Appx:spinEmbedding}. By construction, the ground state space of $\mathcal{H}$ is spanned by assignments that violate the least number of clauses in a given MAX-$2$-SAT instance. 

\section{Logistic Saturation Conjecture}

The state space which can be prepared with a $p$-depth QAOA circuit is defined as,
\begin{equation}
    \Omega = \underset{\boldsymbol{\gamma},\boldsymbol{\beta}}{\bigcup} \ket{\psi(\boldsymbol{\gamma},\boldsymbol{\beta})}, 
\end{equation}
with $\ket{\psi(\boldsymbol{\gamma},\boldsymbol{\beta})}$ given by Eq.~\eqref{qaoa equation}. 
In general, for a given instance size $n$, density $\alpha$ and depth $p$, the variational state space may not cover the whole Hilbert space. Therefore, optimization over $\Omega$ may not converge to the exact ground state energy. The difference, or the energy error in approximation,
\begin{equation}\label{reachabilitydef}
    \text{f}(p,\alpha,n) = \min_{\psi \in \Omega \subseteq \mathbb{C}_2^{\otimes n}} \bra{\psi}\mathcal{H}\ket{\psi} - \min_{\phi \in \mathbb{C}_2^{\otimes n}} \bra{\phi}\mathcal{H}\ket{\phi},
\end{equation} 
quantifies performance of $p$-depth QAOA. Whenever $f(p,\alpha,n) > 0$, QAOA suffers a \textit{reachability deficit}. This behaviour has been studied in \cite{akshay2020reachability,akshay2021reachability}, and it is observed that the onset of these deficits correspond to the phase transition criticality of the considered optimization problem. For the case of MAX-2-SAT, it was shown that $f(p,\alpha,n)$ grows with increasing clause density beyond $ \alpha_c \approx 1$. Though better performance is achieved at the cost of increasing QAOA depth, a similar trend for deficits is reflected.

Motivated by these findings, we propose a \textit{logistic saturation conjecture}:

\noindent Given some fixed tolerance on performance, $\epsilon > 0$. For QAOA on instances characterized by density $\alpha$, the critical depth $p^*$, required to guarantee approximations that fall within an $\epsilon$ tolerance is given by:

\begin{equation} \label{eq:criticaldepthcompute}
    p^*(\alpha) = \min \{p~|~\text{f}(p,\alpha,n) \leq \epsilon\}.
\end{equation}
We conjecture the critical depth to scale with density as,
\begin{equation} \label{eq:criticaldepth}
   p^*(\alpha) \approx \frac{p_{max}}{1+e^{-\kappa(\alpha - \alpha_c)}},
\end{equation} where $\alpha_c$ is the critical density, $\kappa$ the logistic growth rate, and $p_{max}$ the saturation value.    

\section{Results}

\begin{figure}[htp]
\centering
\includegraphics[width=0.47\textwidth]{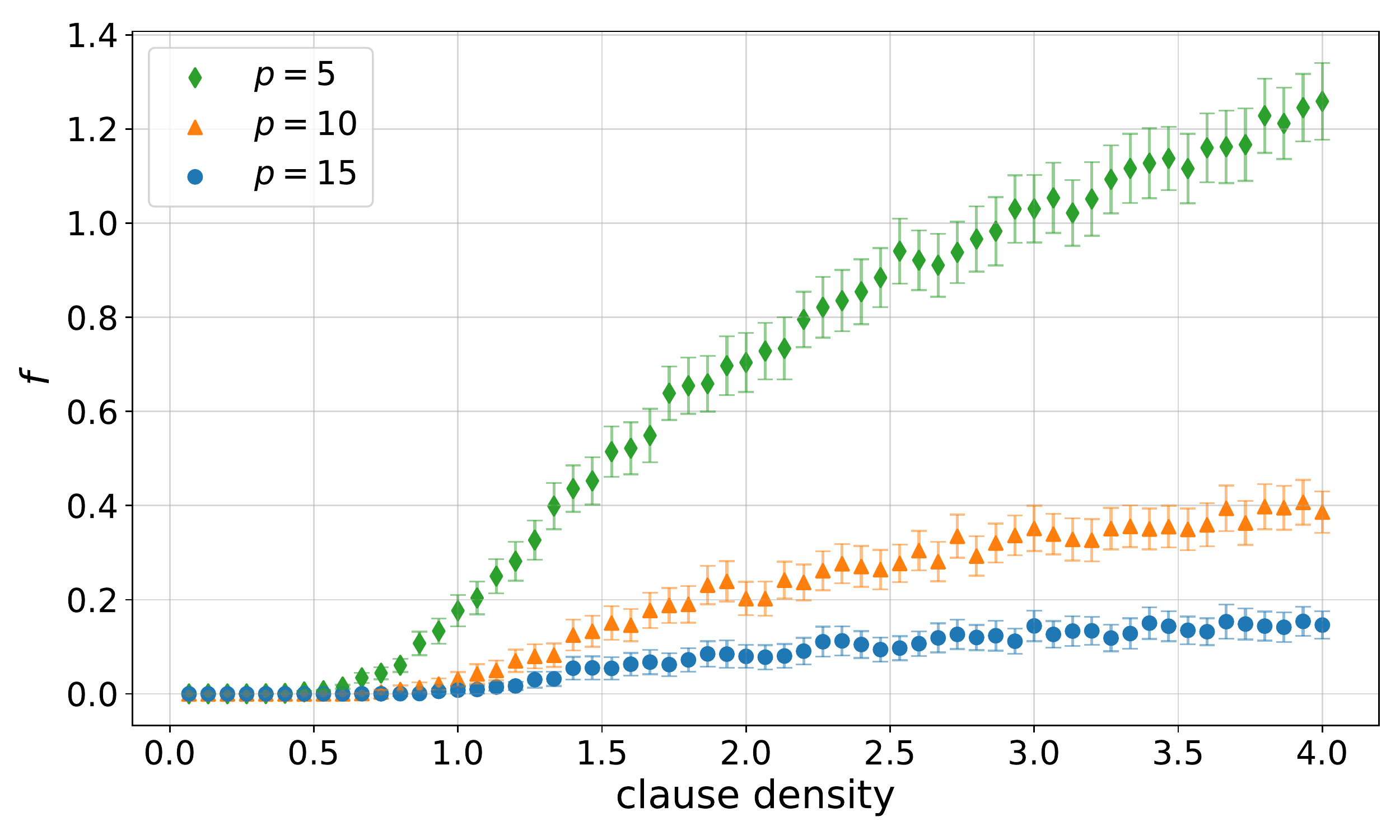}
\caption{Average energy error in approximation, $f$ in Eq.~\eqref{reachabilitydef}, vs clause density for MAX-$2$-SAT instances on $15$ variables. Data points represent the average value obtained over statistics of $300$ random instances with error bars representing $3\sigma$ for the estimated mean. Colors indicate varying QAOA depths $p=\{5,10,15\}$, illustrating improved performance at higher depths.}
\label{QAOAperformance}
\end{figure}

 We first recover the findings of \cite{akshay2020reachability,akshay2021reachability} in Figure~\ref{QAOAperformance} by studying the performance of fixed depth QAOA on $300$ randomly generated MAX-$2$-SAT instances on 15 variables and clauses ranging from 1 to 60 ($\alpha \in \left[0.06, 4.0\right]$).  

\begin{figure}[htb]
    \centering
    \includegraphics[width=0.47\textwidth]{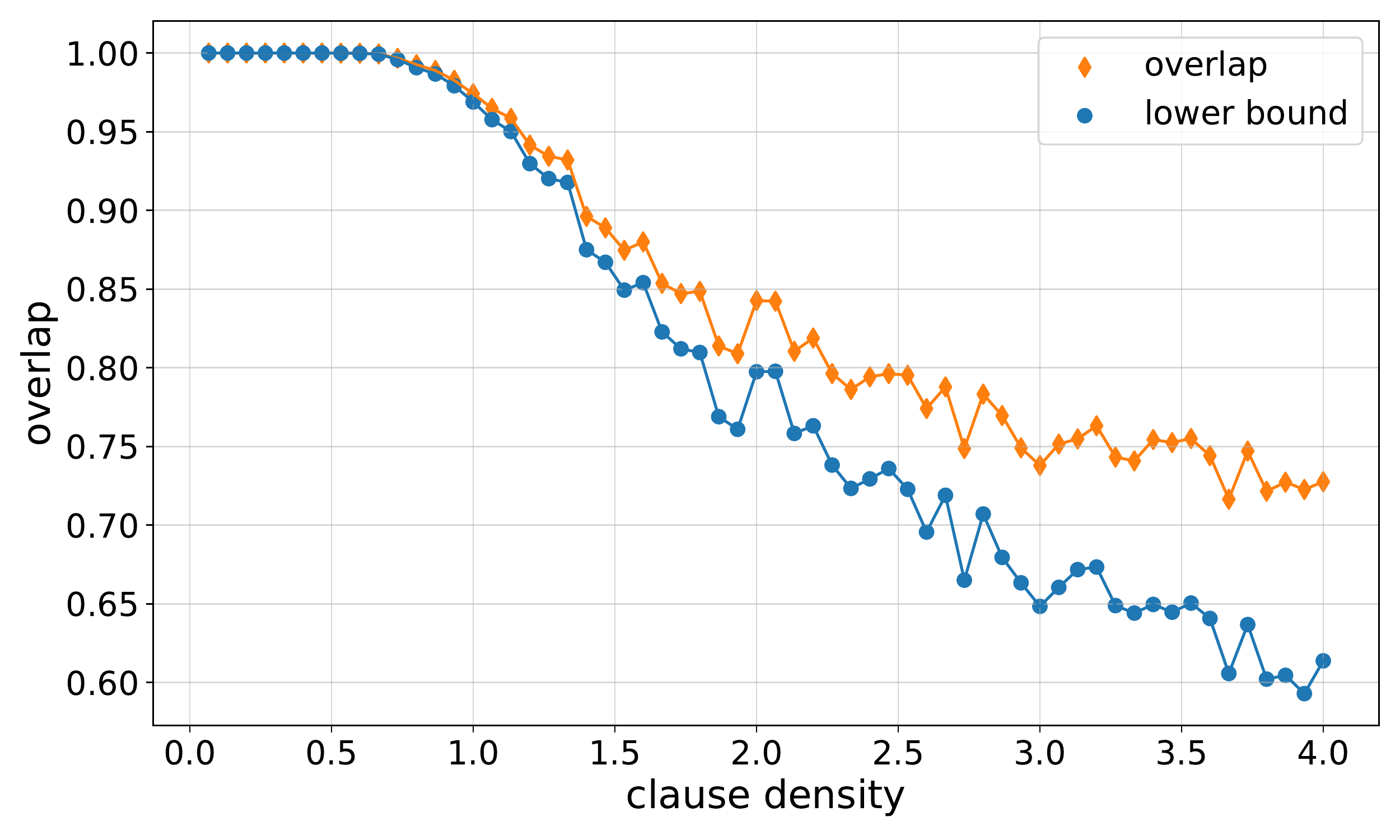}
    \caption{ Ground state overlap vs clause density for depth $p=10$ QAOA on MAX-2-SAT instances with 15 variables. Stability lower bound is calculated by considering average energy error in approximation obtained over statistics of $300$ random instances. Points in orange indicate the corresponding average ground state overlap.}
    \label{fig:sat2_stability}
\end{figure}

\noindent Numerically, it is seen that the average performance worsens monotonically with increasing clause density. This is also evident (see Fig.~\ref{fig:sat2_stability}) for the case of ground state overlap as described by stability bounds in Eq.~\eqref{energy-ovelap}.

\noindent It becomes apparent that for guaranteeing fixed performance, the QAOA depth must necessarily increase with increasing clause density.

\begin{figure}[htp]
\centering
\includegraphics[width=0.47\textwidth]{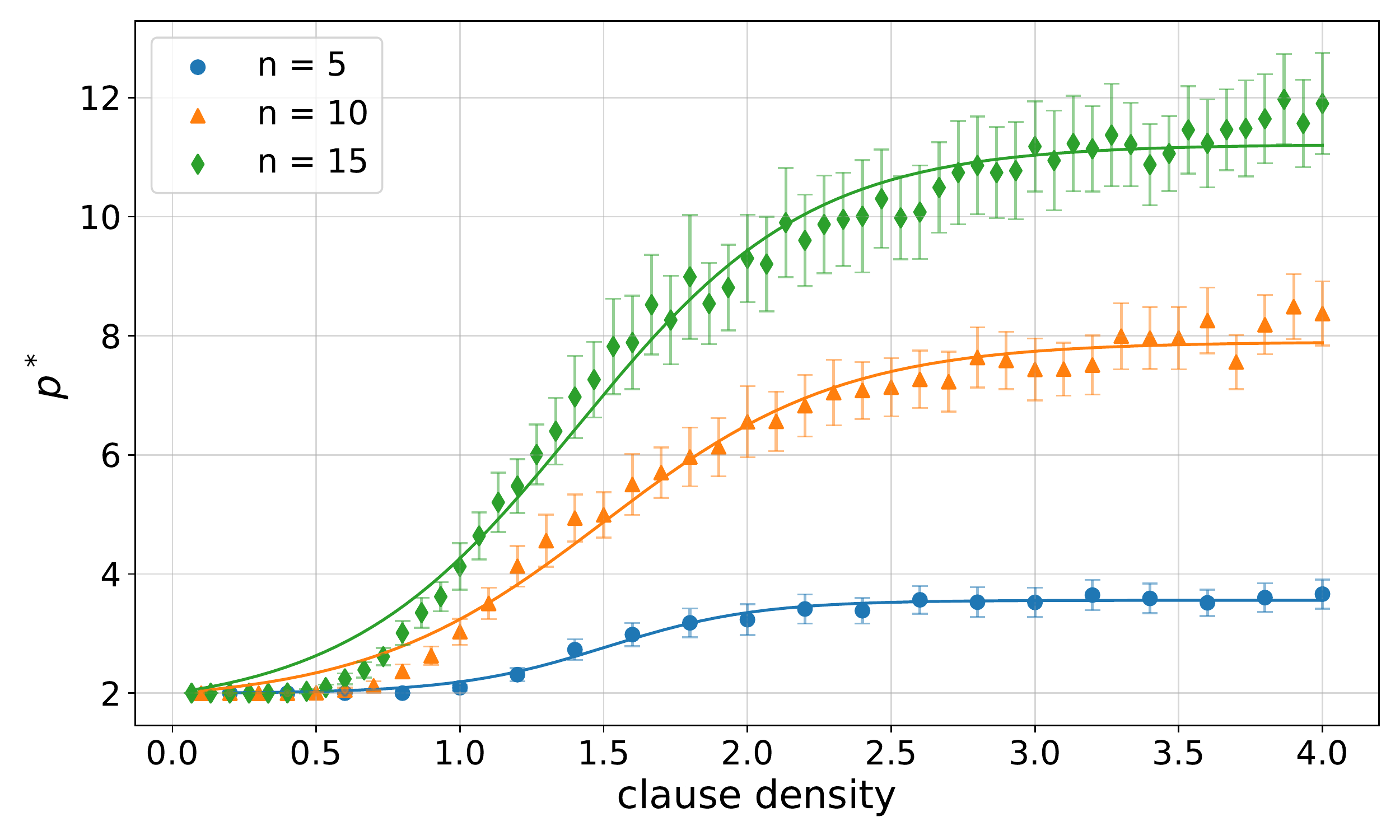}
\caption{Average critical depth $p^*$ vs clause density. Data points represent the average $p^*$ calculated according to Eq. \eqref{eq:criticaldepthcompute} 
across problem density for MAX-$2$-SAT with tolerance $\epsilon = 0.3$. Colors correspond to different problem sizes $n=\{5,10,15\}$ with errorbars representing $3\sigma$ for the estimated mean. Solid curves represent the least squares fit of the data to the predictive model as described in Eq.~\eqref{eq:criticaldepth}. }
\label{QAOAcritical}
\end{figure}

To test the logistic saturation conjecture, we numerically calculate the depth required to attain performance within an $\epsilon$ tolerance. This is done by incrementing depth, until the energy error in approximations falls within acceptance, $~\text{f}(p,\alpha,n) \leq \epsilon$. For each instance this process is repeated and the average critical depth is calculated. Figure~\ref{QAOAcritical} illustrates scaling of the average critical depth with respect to clause density and compares the numerical data against the predictive model in Eq.~\eqref{eq:criticaldepth}. We see that the model is able to match the numerical data well within a $3\sigma$ confidence interval for the estimated mean.

\begin{figure}[htp]
\centering
\includegraphics[width=0.47\textwidth]{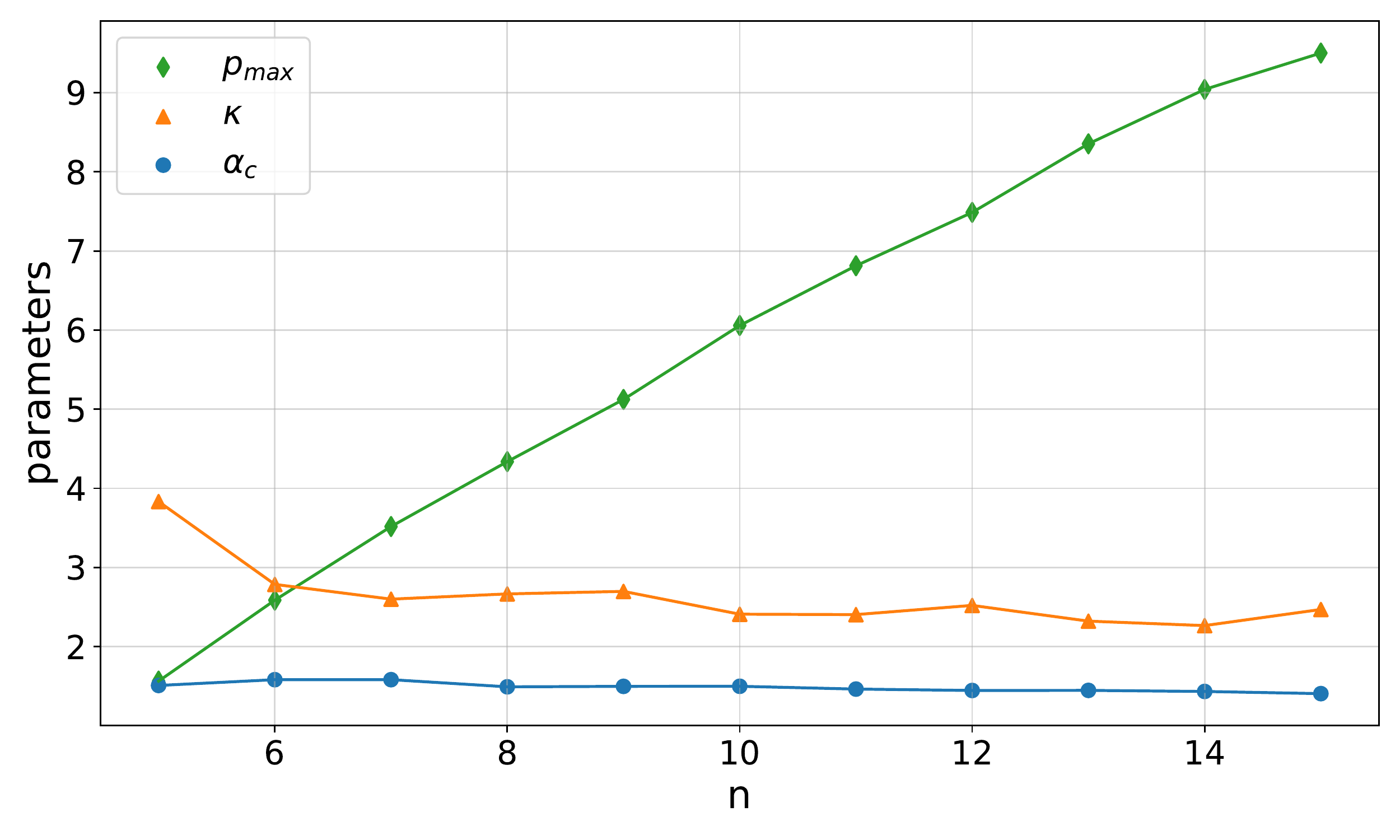}
\caption{Fitting parameters vs number of qubits for QAOA on MAX-$2$-SAT with tolerance $\epsilon = 0.3$. Note that the recovered critical density $\alpha_c \approx 1$, and growth rate $\kappa$, show very little variability with respect to number of qubits. In contrast, the the saturation value $p_{max}$ illustrates a linear trend in the considered range of $5$ to $15$ qubits.}
\label{Logistic_fitting}
\end{figure}

We further investigate the reliability of the predictive model by treating the critical density as a fitting parameter in Eq.~\eqref{eq:criticaldepth} and compare against the theoretical value of $\alpha_c \approx 1$. In Figure~\ref{Logistic_fitting} fitting parameters are recovered for $n = 5$ up to $15$ and indeed we see the critical density recovered from the predictive model matches the theoretical value up to finite size variations.

\section{Numerical Details}

The computationally intensive calculations were performed on the KunLun supercomputer with a maximum CPU frequency of $2.1$ GHz.
The code was written in Python and C\texttt{++}.
Python code is responsible for providing easy-to-use interfaces and C\texttt{++} code is responsible for the actual simulation tasks as the underlying simulation code.
The code executed in parallel with a resource of 1500 cores.
The calculations were performed for 300 random instances for each respective problem density. 

We used a heuristic optimisation strategy, motivated by layerwise training, to find good initial parameterization for the ansatz such that gradient based classical optimizers may avoid most local minima.  

In order to find optimal parameters for depth $p$, we start by optimizing a depth one ansatz (single layer). Then keeping the previous parameters fixed, a new layer is added and trained. This is followed by a simultaneous optimization of all parameters and the process is repeated until the desired depth $p$ is attained.

In our implementation of this heuristic, at each optimization step, the best result out of 25 random seed runs of L-BFGS-B optimizer is chosen.

\section{Conclusion}

QAOA is among the most studied gate based approaches for combinatorial optimization in NISQ era devices. Although analytical techniques address QAOA either at low depth or in the adiabatic limit, $p \to \infty$, understanding performance at intermediate depths remains largely open. Furthermore, problem density induces underparameterization and limits the performance of fixed depth QAOA. An open question in this regard is the required depth for QAOA to succeed. Additionally, by treating the circuit depth of QAOA as a computational resource, it becomes paramount in complexity studies to know the scaling of depth with respect to problem size. Understanding these open problems paves way towards identifying quantum advantage in QAOA.     

In this paper we introduce a methodology based on a predictive model that aims to address these open problems. We first relate QAOA energy approximations and success probability (or ground state overlap) via Eq.~\eqref{energy-ovelap}. This relation then determines success/failure of the algorithm up to some $\epsilon$ tolerance. Based on this acceptance criterion, we numerically recovered the critical depth needed for QAOA to achieve success probability of at least 0.7 for the case of MAX-$2$-SAT. Considering up to $15$ variables and clause density up to 4 we indeed observe that our proposed model describes the data within a $3\sigma$ confidence interval. Furthermore, we recovered for the first time, the scaling of critical depth with respect to problem size which illustrates a linear trend in our considered range of problem sizes. Though this finite range is insufficient to assert quantum advantage, we anticipate future works to test the accuracy of the presented model over comprehensive ranges of densities and problem sizes.

\begin{acknowledgments}

\noindent {\bf Author contributions.} All authors conceived and developed the theory and design of this study and verified the methods. All authors wrote the paper.


\noindent {\bf Data and code availability.} The data that supports the findings of this study are available within the article. The code for generating the data will be available on reasonable request. 

\noindent {\bf Competing interests.} The authors declare no competing interests.
\end{acknowledgments}

\bibliographystyle{unsrt}
\bibliography{ref.bib}

\onecolumngrid
\appendix
\newpage
\section{Ising Embedding of SAT instances} \label{Appx:spinEmbedding}
We start by mapping logical bits $\{0,1\}$ to states $\{\ket{0},\ket{1}\} \in \mathbb{C}_2$ respectively. The objective function to be minimized is equivalent to minimizing a Hamiltonian $\mathcal{H}_{SAT}$, constructed by the real linear extension of $\{P_0, P_1, \openone \}$ as:
\begin{equation}
x_j \longrightarrow P_j^0, \qquad \neg x_j \longrightarrow P_j^1,
\end{equation}
and
\begin{equation}
\wedge \longrightarrow +, \qquad \vee \longrightarrow \otimes.
\end{equation} 
Here $P_j^{1}=\ketbra{1}{1}_j$ and $P_j^{0}=\ketbra{0}{0}_j$ are rank one projectors acting on the $j^{th}$ qubit. The construction ensures each clause in a given SAT instance is mapped onto projectors, $P_{i j \cdots k}^{\alpha \beta \cdots \gamma } = P_{i}^{\alpha} \otimes P_{j}^{\beta} \otimes \cdots \otimes P_{k}^{\gamma}$, which penalize unsatisfiable assignments $\ket{x} \in \mathbb{C}_2^{\otimes n}$ with a penalty $\bra{x} P_{i j \cdots k}^{\alpha \beta \cdots \gamma } \ket{x} = 1$.

\noindent The full Hamiltonian $\mathcal{H}_{SAT}$ is then constructed by summing over all the clauses in an instance,
\begin{equation}\label{eq:sathamiltonian}
    \mathcal{H}_{SAT}=\sum_{l=1}^{m} \mathcal{C}_{l} \lbrace P_{i j...k}^{\alpha \beta...\gamma } \rbrace,
\end{equation} where $\mathcal{C}_{l}$ assigns the value of $\alpha, \beta, \cdots, \gamma \in \{0, 1\}$, corresponding to the negation of the literals indexed by $i, j, \cdots, k \in \{1, 2, \cdots, n\}$ in the $l^{th}$ clause. It is straightforward to see that the ground state space of this SAT-Hamiltonian is spanned by assignments that violate minimal number of clauses.

\noindent Decomposing the projectors onto the $Z$ Pauli basis as,
\begin{equation}
    P^{\alpha}_j=\frac{1}{2}(\openone + (-1)^{\alpha} Z_{j}),    
\end{equation}
  Eq.~\eqref{eq:sathamiltonian} can be re-written in the form of a generalized Ising Hamiltonian with at most k-body interaction. In case of 2-SAT, the Hamiltonian $\mathcal{H}_{2SAT}$, contain at most 2-body interactions as,
\begin{equation}
    \begin{split}
         \mathcal{H}_{2SAT} &= \sum_{l=1}^{m} \mathcal{C}_l \lbrace P^{\alpha}_j \otimes P^{\beta}_k\rbrace\\
    &= \frac{1}{4} \left( m \openone + \sum_{j}h_{j} Z{j}   +  \sum_{j < k}\mathcal{J}_{jk} Z_{j}Z_{k}\right),
    \end{split}
\end{equation}
with appropriate coefficients $h_{j}$ and $\mathcal{J}_{jk}$.

\section{Derivation of Energy-Overlap Bounds.}\label{Appx:stability}
Once a variatinal state $\ket{\psi(\boldsymbol{\theta})}$ has been prepared on a quantum computer as a function of a set of tunable parameters $\boldsymbol{\theta}$, we would like to know how close this state is to the solution space of the problem being solved. We will adapt and extend the Lemma 1 (Variational Stability) from \cite{biamonte2021universal} to the degenerate case. 

Let $ \mathcal{H} \geq 0$ be some Hamiltonian with $d$--degenerate ground state energy $\lambda_{0} \geq 0$. The Hamiltonian in its eigenbasis can be expressed as,
\begin{equation}
    \mathcal{H} =  \lambda_0 \sum_{j=1}^{d}  \ketbra{\phi_j}{\phi_j} + \sum_{k > d} \lambda_k  \ketbra{\phi_k}{\phi_k}.
\end{equation}
Here we reorder the eigenbasis so that $\lambda_0 < \lambda_1 \leq \cdots \lambda_k \cdots  \leq \lambda_{max}$.  
Consider the expectation value of the Hamiltonian on an arbitrary state $\ket{\psi}$,

\begin{equation}\label{expected}
     \bra{\psi}\mathcal{H}\ket{\psi} =  \lambda_0 \sum_{j=1}^{d}  \abs{\braket{\phi_{j}}{\psi}}^{2} + \sum_{k > d} \lambda_k \abs{\braket{\phi_{k}}{\psi}}^{2}.
\end{equation}

Assuming the gap $\Delta = \lambda_1 - \lambda_0$, is known, 
\begin{equation*}\label{ineq1}
\begin{split}
    \bra{\psi}\mathcal{H}\ket{\psi} &= \lambda_0 \sum_{j=1}^{d} \abs{\braket{\phi_{j}}{\psi}}^{2} + \sum_{k > d} \lambda_k \abs{\braket{\phi_{k}}{\psi}}^{2} \\
    & \geq \lambda_0 (\sum_{j=1}^{d} \abs{\braket{\phi_{j}}{\psi}}^{2}) + (\lambda_0 +\Delta)(\sum_{k > d} \abs{\braket{\phi_{k}}{\psi}}^{2}) \\
    & = \lambda_0 (\sum_{j=1}^{d} \abs{\braket{\phi_{j}}{\psi}}^{2}) + (\lambda_0 +\Delta)(1-\sum_{j=1}^{d} \abs{\braket{\phi_{j}}{\psi}}^{2})\\
    &= \lambda_0 +\Delta(1-\sum_{j=1}^{d} \abs{\braket{\phi_{j}}{\psi}}^{2}).
\end{split}
\end{equation*}
Rearranging the terms then gives the required lower bound on the ground state as: 
\begin{equation}
    \sum_{j=1}^{d} \abs{\braket{\phi_{j}}{\psi}}^{2} \geq 1 - \frac{\bra{\psi}\mathcal{H}\ket{\psi}-\lambda_0}{\Delta}.
\end{equation}

For the upper bound, we substitute for each excited state energies $\lambda_k$ in Eq.~\eqref{expected} by $ \lambda_0 + \lambda_{max}$,
\begin{equation}
\begin{split}
    \bra{\psi}\mathcal{H}\ket{\psi} &\leq   \lambda_0 \sum_{j=1}^{d} \abs{\braket{\phi_{j}}{\psi}}^{2} + (\lambda_0 +\lambda_{max}) \sum_{k>d} \abs{\braket{\phi_{k}}{\psi}}^{2} \\
    & \leq \lambda_0 (\sum_{j=1}^{d} \abs{\braket{\phi_{j}}{\psi}}^{2}) \\& ~~~~ + (\lambda_0 +\lambda_{max}) (1 - \sum_{j=1}^{d} \abs{\braket{\phi_{j}}{\psi}}^{2}). \\
\end{split}
\end{equation}
Rearranging the terms now gives the required upper bound as:  
\begin{equation}
    \sum_{j=1}^{d} \abs{\braket{\phi_{j}}{\psi}}^{2} \leq 1 - \frac{\bra{\psi}\mathcal{H}\ket{\psi} - \lambda_0}{\lambda_{max}}.
\end{equation}

By combining both bounds, we obtain the stability theorem that relates energy error in approximation with the success probability (or the ground state overlap) for the degenerated case: 
\begin{equation}
    1 - \frac{f}{\Delta} \leq  \sum_{i=1}^{d} \abs{\braket{\phi_{i}}{\psi}}^{2} \leq 1 - \frac{f}{\lambda_{max}},
\end{equation}
where $f = \bra{\psi}\mathcal{H}\ket{\psi} - \lambda_0$. 

\noindent  Notice that for the case of MAX-$2$-SAT instances, $\lambda_{max} \leq m$ where $m$ the number of clauses in the instance and $\Delta \geq 1$.  

\end{document}